\documentstyle[12pt]{article}
\input epsf.tex

\def\Z{{\bf Z}}
\def\N{{\rm N}}
\def\S{{\bf S}}
\def\R{{\bf R}}

\def\TL{\hfil$\displaystyle{##}$}
\def\TR{$\displaystyle{{}##}$\hfil}
\def\TC{\hfil$\displaystyle{##}$\hfil}

\def\displaylines#1{\vcenter{\openup1\jot
  \halign{\strut\span\TC\cr #1 \cr}}}

\def\comment#1{}
\def\fixit#1{}

\def\tf#1#2{{\textstyle{#1 \over #2}}}

\def\Tr{\mop{Tr}}

\def\href#1#2{#2}  

\def\lbldef#1#2{\expandafter\gdef\csname #1\endcsname {#2}}
\def\eqn#1#2{\lbldef{#1}{(\ref{#1})}%
\begin{equation} #2 \label{#1} \end{equation}}
\def\eqalign#1{\vcenter{\openup1\jot
    \halign{\strut\span\TL & \span\TR\cr #1 \cr
   }}}
\def\eno#1{(\ref{#1})}

\begin{document}

\baselineskip=15.5pt
\pagestyle{plain}
\setcounter{page}{1}

\renewcommand{\thefootnote}{\fnsymbol{footnote}}

\begin{titlepage}

\begin{flushright}
PUPT-1807\\
hep-th/9808075
\end{flushright}
\vfil

\begin{center}
{\huge Baryons and Domain Walls in an 
${\cal N} = 1$ Superconformal Gauge Theory
}
\end{center}

\vfil
\begin{center}
{\large Steven S.\ Gubser and
Igor R.\ Klebanov}\\
\vspace{1mm}
Joseph Henry Laboratories\\
Princeton University\\
Princeton, New Jersey 08542, USA\\
\vspace{3mm}
\end{center}

\vfil

\begin{center}
{\large Abstract}
\end{center}

\noindent
Coincident D3-branes placed at
a conical singularity are related to string theory on $AdS_5\times X_5$,
for a suitable five-dimensional Einstein manifold $X_5$.  
For the example of the conifold, which leads to
$X_5=T^{1,1}=(SU(2)\times SU(2))/U(1)$, the infrared limit of
the theory on $N$ D3-branes was constructed recently.
This is ${\cal N}=1$ supersymmetric $SU(N)\times SU(N)$ gauge theory
coupled to four bifundamental chiral superfields 
and supplemented by a quartic superpotential which becomes
marginal in the infrared.
In this paper we consider D3-branes wrapped over the 3-cycles of
$T^{1,1}$ and identify them with baryon-like chiral operators built
out of products of $N$ chiral superfields. The supergravity calculation of
the dimensions of such operators agrees with field theory.
We also study the D5-brane 
wrapped over a 2-cycle of $T^{1,1}$, which acts as a domain wall
in $AdS_5$. We argue that upon crossing it the gauge group changes
to $SU(N)\times SU(N+1)$. This suggests a construction of
supergravity duals of ${\cal N}=1$ supersymmetric 
$SU(N_1)\times SU(N_2)$ gauge theories.

\vfil
\begin{flushleft}
August 1998
\end{flushleft}
\end{titlepage}
\newpage

\renewcommand{\thefootnote}{\arabic{footnote}}
\setcounter{footnote}{0}

\def\R{{\bf R}}
\def\RP{{\bf RP}}
\def\Tr{{\rm Tr}}
\def\N{{\rm N}}

\newcommand{\grad}{\nabla}
\newcommand{\half}{{1\over 2}}
\newcommand{\third}{{1\over 3}}
\newcommand{\be}{\begin{equation}}
\newcommand{\ee}{\end{equation}}
\newcommand{\bea}{\begin{eqnarray}}
\newcommand{\eea}{\end{eqnarray}}

\newcommand{\dint}[2]{\int\limits_{#1}^{#2}}
\newcommand{\D}{\displaystyle}
\newcommand{\PDT}[1]{\frac{\partial #1}{\partial t}}
\newcommand{\PD}{\partial}
\newcommand{\tw}{\tilde{w}}
\newcommand{\tg}{\tilde{g}}
\newcommand{\newcaption}[1]{\centerline{\parbox{6in}{\caption{#1}}}}
\def\href#1#2{#2}  

\def \ci {\cite}
\def \foot {\footnote}
\def \bi{\bibitem}
\newcommand{\rf}[1]{(\ref{#1})}
\def \del{\partial}
\def \m {\mu}
\def \n {\nu} 
\def \g {\gamma}
\def \G {\Gamma}
\def \a {\alpha}
\def \ov {\over}
\def \la {\label}
\def \ep {\epsilon}
\def \d {\delta}
\def \k {\kappa}
\def \p {\phi}
\def \ha {\textstyle{1\ov 2}}

\def\np {  {\em Nucl. Phys.} }
\def \pl { {\em Phys. Lett.} }
\def \mpl { Mod. Phys. Lett. }
\def \prl { Phys. Rev. Lett. }
\def \pr  { {\em Phys. Rev.} }
\def \cqg { Class. Quantum Grav.}
\def \jmp { Journ. Math. Phys. }
\def\ap { Ann. Phys. }
\def \ijmp { Int. J. Mod. Phys. }

\section{Introduction}
\label{Intro}

Over two decades ago 't~Hooft showed that gauge theories simplify
in the limit where the number of colors, $N$, is taken to infinity
\cite{GT}. A number of arguments suggest that, for large $N$,
gauge theories have a dual description in terms of string theory
\cite{GT,Sasha}. Recently, with some motivation from the D-brane
description of black three-branes \cite{gkp,kleb,gkThree},
and from studies of the throat geometry \cite{gt}, 
Maldacena argued \cite{jthroat} that
${\cal N}=4$ supersymmetric $SU(N)$ gauge theory is related to Type IIB
strings on $AdS_5\times \S^5$.\foot{
This Type IIB background was originally considered in \cite{JH}.}
This correspondence was sharpened in 
\cite{US,EW},
where it was shown how to calculate the correlation functions of  
gauge theory operators from the response of the Type IIB theory
on $AdS_5\times \S^5$ to boundary conditions. 

According to general arguments presented in \cite{EW},
Type IIB theory on $AdS_5\times X_5$, where $X_5$ is a
five-dimensional Einstein manifold bearing five-form flux, 
is expected to be dual to a four-dimensional
conformal field theory. Construction of field theories 
for various manifolds $X_5$, in addition to the maximally supersymmetric
case $X_5= \S^5$, has become an active area.
In one class of examples, $X_5$ is an
$\S^5$ divided by the action of a discrete group. The field theory
one thus obtains is the infrared limit of the world volume 
theory on $N$ D3-branes \cite{brane,Witten}
placed at an orbifold \cite{dm,ks,lnv,bkv}
or 7-brane and orientifold singularity \cite{k,fs,afm}. 

Very recently, a new example of duality
was found \cite{KW}
where $X_5$ is a smooth Einstein manifold whose local geometry
is different from that of $\S^5$. The $X_5$ for which the
dual field theory was constructed in 
\cite{KW} is one of the
coset spaces $T^{p,q}= (SU(2)\times SU(2))/U(1)$
originally considered by Romans in the context
of Kaluza-Klein supergravity \cite{Romans}.
The $U(1)$
is a diagonal subgroup of the maximal torus of $SU(2)\times SU(2)$:
if $\sigma^{L,R}_i$ are the generators of the left and right
$SU(2)$'s, then the $U(1)$  is generated by
$p\sigma^L_3 + q\sigma^R_3$. Romans found that for $p=q=1$ the 
compactification preserves 8 supersymmetries, while for other $p$
and $q$ all supersymmetries are broken. Therefore, one expects
type IIB theory on $AdS_5\times T^{1,1}$ to be dual to a certain
large $N$ ${\cal N}=1$ superconformal field theory in four dimensions.
The dual theory constructed in \cite{KW} turns out to be a non-trivial
infrared fixed point. It is the $SU(N)\times SU(N)$
gauge theory with 
chiral  superfields $A_k, \,k=1,2$ transforming in the
$(\N,\overline{\N})$ representation and $B_l,\,l=1,2$ 
transforming in the
$(\overline{\N},\N)$ representation. 
These fields acquire infrared anomalous dimensions equal to $-1/4$
determined by the existence of an anomaly-free R-symmetry.
A crucial ingredient in the construction of \cite{KW} is the quartic 
superpotential $W=\lambda\Tr (A_1B_1A_2B_2-A_1B_2A_2B_1)$ which becomes
exactly marginal in the infrared. 

The construction of the field theory in \cite{KW}
was guided by the observation that it is the infrared limit
of the world volume theory on coincident 
Dirichlet three-branes 
placed at a conical singularity of
a non-compact Calabi-Yau (CY) threefold (this is
a special case
of the connection between compactification on Einstein manifolds
and the metric of three-branes placed at
conical singularities \cite{Kehagias,KW,Morr,ff}).
The CY manifold relevant here is the simplest non-compact 
threefold with a conical singularity. This is the conifold 
\cite{cgh,cd},
which for our purposes is the complex manifold $C$
\be\label{iconifold}
z_1^2+z_2^2+z_3^2+z_4^2=0
\ee
with a ``double point'' singularity at $z_i=0$.
The metric on the conifold may be written as
\be
\label{metric}
ds_6^2 = dr^2 + r^2 g_{ij}(x) dx^i dx^j\ , \qquad (i,j= 1, \ldots, 5)
\ .
\ee
Here $g_{ij}$ is the metric on the base of the cone, which is precisely
the Romans manifold $T^{1,1}$ \cite{cd,KW}. The isometries of $T^{1,1}$,
which form the group $SU(2)\times SU(2)\times U(1)$, are realized
very simply in terms of the $z$-coordinates.
The $z_k$ transform in the four-dimensional representation of 
$SO(4)= SU(2)\times SU(2)$, while under the $U(1)$ $z_k\to e^{i\alpha}
z_k$. The metric on $T^{1,1}$ may be written down explicitly by
utilizing the fact that it is a
$U(1)$ bundle over $\S^2\times \S^2$.
Choosing the coordinates $(\theta_1,\phi_1)$ and $(\theta_2,\phi_2)$
to parametrize the two spheres in a conventional way, and
the angle $\psi \in [0, 4\pi)$ to parametrize the $U(1)$  fiber,
the metric may be written as \cite{cd}
\be
\label{tmetric}
g_{ij}(x) dx^i dx^j = {1\over 9}
(d\psi + \cos \theta_1 d\phi_1+ \cos \theta_2 d\phi_2)^2+
{1\over 6} \sum_{i=1}^2 \left [
d\theta_i^2 + \sin^2\theta_i d\phi_i^2 \right ]
\ .
\ee
Explicit calculation shows that this metric is indeed Einstein,
$R_{ij} = 4 g_{ij}$ \cite{cd,Page}.

It turns out that the coset space 
$T^{1,1}=(SU(2)\times SU(2))/U(1)$
may be obtained by 
blowing up the fixed circle of $\S^5/{\bf Z}_2$ (an operation that
breaks ${\cal N}=2$ to ${\cal N}=1$) \cite{KW}.
On the field theory side, the ${\cal N}=2$ superconformal theory
corresponding to $\S^5/{\bf Z}_2$ flows to the ${\cal N}=1$
IR fixed point corresponding to $T^{1,1}$. The necessary relevant
perturbation of the superpotential is odd under the
${\bf Z}_2$ and therefore
corresponds to a blow-up mode of the orbifold \cite{Han,Guk}.
It is interesting to examine how various quantities change under
the RG flow from the $\S^5/{\bf Z}_2$ theory to the
$T^{1,1}$ theory. The behavior of the conformal anomaly (which is equal to
the $U(1)^3_R$ anomaly) was studied in \cite{SG}. Using the values of the
R-charges deduced in \cite{KW}, on the field theory
side it was found that 
\be \label{anom}
{c_{IR}\over c_{UV}} = {27\over 32}
\ .
\ee
On the other hand, all 3-point functions calculated from supergravity on
$AdS_5\times X_5$ carry normalization factor inversely proportional
to ${\rm Vol}\ (X_5)$. Thus, on the supergravity side
\be
{c_{IR}\over c_{UV}} = {{\rm Vol}\ (\S^5/{\bf Z}_2)
\over {\rm Vol}\ (T^{1,1}) }
\ .
\ee
Since \cite{SG} 
\be 
\label{volume}
{\rm Vol}\ (T^{1,1})= {16 \pi^3\over 27} \ , 
\qquad {\rm Vol}\ (\S^5/{\bf Z}_2)= {\pi^3\over 2}\ ,
\ee
the supergravity calculation
is in exact agreement with the field theory result (\ref{anom}).
This is a striking and highly sensitive test of the ${\cal N}=1$
dual pair constructed in \cite{KW}.

In this letter we carry out further studies of this dual pair.  In
particular, we consider various branes wrapped over the cycles of
$T^{1,1}$ and attempt to 
identify these states in the field theory. Wrapped
D3-branes turn out to correspond to baryon-like operators $A^N$
and $B^N$ where the indices of both $SU(N)$ groups are fully
antisymmetrized.  For large $N$ the dimensions of such operators
calculated from the supergravity are found to be $3N/4$. This is in
complete agreement with the fact that the dimension of the chiral
superfields at the fixed point is $3/4$ and may be regarded as a direct
supergravity calculation of an anomalous dimension in the dual gauge
theory.

We further argue
that a domain wall made out of a D5-brane wrapped over a 2-cycle of 
$T^{1,1}$ separates a $SU(N)\times SU(N)$ gauge theory from
a $SU(N)\times SU(N+1)$ gauge theory. Indeed, passing a wrapped D3-brane
through such a domain wall produces a fundamental string stretched
between the D3-brane and the domain wall. On the 
$SU(N)\times SU(N+1)$ side
we find baryon-like operators which transform in the fundamental
representation of one of the gauge groups, and identify them with
the wrapped D3-brane attached to a string.

\section{Three-cycles and Baryon-like Operators}
\label{Baryons}

After placing a large number $N$ of coincident D3-branes at
the singularity of the conifold and taking the near-horizon limit,
the metric becomes that of $AdS_5\times X_5$,
\be
ds_{10}^2= {r^2\over L^2} \eta_{\mu\nu} dy^\mu dy^\nu
+ L^2 \left ({dr^2\over r^2} + g_{ij}(x) dx^i dx^j\right )
\ .
\ee
The scale $L$ is related to $N$ and the gravitational constant $\kappa$
through \cite{SG}\footnote{An easy way to derive this relation
is by equating, as in \cite{gkp}, the ADM tension of the three-brane solution, 
$2 {\rm Vol}\ (X_5) L^4/\kappa^2$, to the tension of $N$ D3-branes,
$N\sqrt\pi/\kappa$.}
\be
\label{scalrel}
L^4 = {\sqrt \pi \kappa N\over 2 {\rm Vol}\ (X_5)}
\ .
\ee  

Now consider wrapping a D3-brane over a 3-cycle of $T^{1,1}$. Topologically,
$T^{1,1}$ is $\S^2\times \S^3$ which establishes the existence of a 
3-cycle \cite{cd}.
In fact, this is a supersymmetric cycle; this fact was
used in \cite{strom} where a three-brane wrapped over this cycle 
was argued to give rise to a massless black hole.
An immediate guess for a 3-cycle of minimum volume is to consider
the subspace at a constant value of $(\theta_2, \phi_2)$ in the metric
(\ref{tmetric}). 
We have checked that the three-brane
wrapped over $(\psi,\theta_1,\phi_1)$ coordinates indeed satisfies
the equations of motion and is thus a minimum volume 
configuration.\footnote{Equivalently, we could consider the subspaces at
constant $(\theta_1, \phi_1)$ and the significance of this in the dual
field theory will become clear shortly.} To calculate the 3-volume,
we need to find the determinant of the following metric:
\be 
\label{threemetric}
{L^2\over 9}
(d\psi + \cos \theta d\phi)^2+
{L^2\over 6} \left [
d\theta^2 + \sin^2\theta d\phi^2 \right ]
\ .
\ee
 Integrating the square root of the determinant over the three
coordinates, we find $V_3= 8\pi^2 L^3/9$.  The mass of the three-brane
wrapped over the 3-cycle is, therefore,
 \be
m= V_3 {\sqrt\pi\over \kappa} = {8 \pi^{5/2} L^3\over 9\kappa}
\ .
\ee
To relate this to the dimension $\Delta$
of the corresponding operator in the
dual field theory, we use the results of \cite{US,EW} 
which for large $m L$ imply $\Delta = mL$. Using (\ref{scalrel})
and (\ref{volume}) for the case of $T^{1,1}$, we find
\be 
\Delta = m L= 
{8 \pi^{5/2} L^4\over 9\kappa} = {3\over 4} N
\ .
\ee

What are the operators in the dual field theory whose dimensions grow as
$N$? The answer is clear: since the fields $A^{\alpha}_{k\beta}$
carry an index $\alpha$ in the $\N$ of $SU(N)_1$ and an index $\beta$
in the $\overline{\N}$ of $SU(N)_2$,
we can construct a baryon-like color-singlet operator
by antisymmetrizing completely with respect to both groups.
The resulting operator has the form
\be\label{BaryonOne}
{\cal B}_{1 l}= \epsilon_{\alpha_1 \ldots \alpha_N}  
\epsilon^{\beta_1\ldots \beta_N} D_{l}^{k_1\ldots k_N}  
\prod_{i=1}^N A^{\alpha_i}_{k_i  \beta_i} 
\ ,
\ee
where $D_l^{k_1\ldots k_N}$ is the completely 
symmetric $SU(2)$ Clebsch-Gordon
coefficient corresponding to forming the $N+1$ of $SU(2)$ out of $N$ 2's.
Thus the $SU(2)\times SU(2)$ quantum numbers of $B_{1 l}$ are
$(N+1, 1)$. Similarly, we can construct baryon-like operators
which transform as $(1, N+1)$,
\be\label{BaryonTwo}
{\cal B}_{2 l}= \epsilon^{\alpha_1 \ldots \alpha_N}  
\epsilon_{\beta_1\ldots \beta_N} D_{l}^{k_1\ldots k_N}  
\prod_{i=1}^N B^{\beta_i}_{k_i  \alpha_i} 
\ .
\ee
Under the duality these operators map to D3-branes classically localized at 
a constant $(\theta_1,\phi_1)$. Thus, the existence of two types
of baryon operators is related on the supergravity
side to the fact that the base of the $U(1)$
bundle is $\S^2\times \S^2$.    

We can further explain why one of the $SU(2)$ quantum numbers is
precisely $N+1$. As shown in \cite{Ed} in 
the context of an analogous construction of Pfaffian operators,
it is necessary to carry out collective coordinate quantization
of the wrapped D3-brane.
While classically the wrapped D3-brane is localized at a point in the 
remaining two dimensions, quantum mechanically we have to find its 
collective coordinate wave
function. In the present case the wrapped D3-brane acts as a charged
particle, while the 5-form field flux through $T^{1,1}$
effectively gives rise to ordinary magnetic flux
through the $\S^2$. We need to ask how many different ground
states there are for a charged particle on a sphere with $N$ units of
magnetic flux. The answer to this problem is well-known: $N+1$.
This degeneracy is due to the fact that the
lowest possible angular momentum of a non-relativistic 
charged particle in the field of a monopole carrying $N$
elementary units of magnetic charge is $N/2$ \cite{SC}.
Thus, the ground state collective coordinate wave functions form
an $N+1$-dimensional representation of the $SU(2)$ that rotates the
$\S^2$ which is not wrapped by the D3-brane (the D3-brane is
obviously a singlet under the other $SU(2)$).
The infinity of classical ground states is
turned into $N+1$ quantum mechanical ground states. The $SU(2)\times
SU(2)$ quantum numbers of the collective-coordinate quantized wrapped
D3-branes are exactly the same as those of the baryon-like operators
(\ref{BaryonOne}), (\ref{BaryonTwo}). This can be regarded as a new test
of the AdS/CFT duality at finite $N$.

Finally, let us compare the actual dimensions of the
operators.
Since the $A$'s and the $B$'s have infrared dimension 3/4 in the construction
of \cite{KW}, we see that the dimension of the baryon-like operator
is indeed $3N/4$, in perfect agreement with supergravity.\footnote{It
is possible that there are $1/N$ corrections to the field theory
result which would be difficult to see in supergravity.}
We regard this as a highly non-trivial check of both
the AdS/CFT correspondence and of the construction of the dual ${\cal N}=1$
superconformal field theory in \cite{KW}. 
 
As a slight digression, and also to check the consistency of our
approach, we show following \cite{Ed} that an analogous calculation
with a wrapped D3-brane produces agreement with the field theoretic
dimension of the Pfaffian operator in $SO(2N)$ gauge theory,
\be \label{Pfaff}
 \epsilon_{a_1 \ldots a_{2N}} \Phi^{a_1 a_2} \ldots
\Phi^{a_{2N-1} a_{2N} } 
\ .
\ee
Since the dimension of $\Phi$ is not renormalized in this case,
we see that the dimension of the Pfaffian operator is equal to $N$.

The $SO(2N)$ theory is dual to supergravity on $AdS_5\times \RP^5$,
and now
\be
\label{newscalrel}
L^4 = {\sqrt \pi \kappa N\over 2 {\rm Vol}\ (\RP^5)}= 
{\kappa N\over \pi^{5/2}} 
\ .
\ee  
The object dual to the Pfaffian operator is
the D3-brane wrapped over $\RP^3= \S^3/{\bf Z}_2$, 
whose volume is $V_3= \pi^2 L^3$. Thus,
$$\Delta = L V_3{\sqrt \pi\over \kappa}= N
\ ,
$$ 
once again in perfect agreement with the field theory.

In many orbifold theories \cite{ks} there are analogues closer than
the Pfaffian of the $SO(2N)$ theory to the baryons considered in
(\ref{BaryonOne}) and (\ref{BaryonTwo}).  Namely, from a bifundamental
matter field $A$ charged under two gauge groups of the same size, one
can make a singlet operator by completely antisymmetrizing both upper
and lower color indices.\footnote{There are more exotic possibilities
involving $k_2$ lower index $\epsilon$-tensors and $k_1$ upper index
$\epsilon$-tensors with $k_1 k_2 N$ powers of a field $A$ in a $({\rm
k_1 N},\overline{\rm k_2 N})$ of $SU(k_1 N) \times SU(k_2 N)$ 
where $k_1$ and $k_2$ are relatively prime.}
We will mention the two simplest examples.  An ${\cal
N}=1$ theory results from the transitive $\Z_3$ orbifold action on
$\S^5$ defined by coordinatizing ${\bf R}^6$ by three complex numbers
$z_1$, $z_2$, $z_3$ and considering the map $z_k \to e^{2\pi i / 3}
z_k$ for all $k$.  The theory has gauge group $SU(N)^3$ with three
$(\N,\overline{\N})$ representations between each pair of gauge groups.
Baryons formed as in (\ref{BaryonOne}) and (\ref{BaryonTwo}) from the
bifundamental matter have dimension $N$.  Minimal area 3-cycles on
$\S^5/{\bf Z}_3$ can be constructed by intersecting the 4-plane
$z_k=0$ for any particular $k$ with the sphere $|z_1|^2 + |z_2|^2 +
|z_3|^2 = 1$.  Now we have 
\be
L^4 = {3 \over 2} {\kappa N \over \pi^{5/2}}\ ,\qquad
V_3 = {2 \over 3} \pi^2 L^3\ , 
\ee
and we find that $\Delta = L V_3{\sqrt \pi\over \kappa}= N$ as
expected.  

As a second example we may consider the ${\cal N}=2$ $\S^5 / \Z_2$
theory.  In this case the orbifold group does not act freely, but has
a circle of fixed points on $\S^5$.  The blowup of the orbifold can be
depicted as an $\S^3$ bundle over $\S^2$ \cite{KW}.  The $\S^3$ fibers
in this bundle would be three-dimensional analogs of great circles
on $\S^5$, except that the $\Z_2$ acts on them by identifying a point
with its image under a $180^\circ$ rotation.  Their volume is thus cut
in half, and for a D3-brane wrapping a fiber we have
 \be 
L^4 = {\kappa N \over \pi^{5/2}}\ ,\qquad 
V_3 = \pi^2 L^3\ .
\ee 
 Once again, $\Delta = L V_3{\sqrt \pi\over \kappa} = N$ in
agreement with the field theory.

\section{Domain Walls in $AdS_5$}
\label{Domain}

Domain walls in a holographic theory come from three-branes 
in $AdS_5$ \cite{Ed}.
The simplest example is a
D3-brane which is not wrapped over the compact manifold.  
Through an analysis of the five-form flux carried over
directly from \cite{Ed} one can conclude that when one crosses the
domain wall, the effect in field theory is to change the gauge group
from $SU(N) \times SU(N)$ to $SU(N+1) \times SU(N+1)$.

The field theory interpretation of a D5-brane wrapped around $\S^2$ is
less obvious.  Recall that $T^{1,1}$ has the topology of $\S^3 \times
\S^2$, so there is topologically only one way to wrap the D5-brane.  If
on one side of the domain wall we have the original $SU(N) \times
SU(N)$ theory, then we claim that on the other side the theory is
$SU(N) \times SU(N+1)$.\foot{We are grateful to O.~Aharony for
useful discussions of this possibility.}  
The matter fields $A_k$ and $B_k$ are still
bifundamentals, filling out $2 (\N,\overline{\N+1}) \oplus 2
(\overline{\N}, \N+1)$.  
An anti-D5-brane wrapped around $\S^2$ will act as a
domain wall which decrements the rank of one gauge group, so that
traversing a D5 and then an anti-D5 leads one back to the original
$SU(N) \times SU(N)$ theory.

The immediate evidence for this claim is the way the baryons
considered in section~\ref{Baryons} behave when crossing the D5-brane
domain wall.  In homology there is only one $\S^3$, but for
definiteness let us wrap the D3-brane around a particular three-sphere
$\S^3_{(1)}$ which is invariant under the group $SU(2)_B$ under which
the fields $B_k$ transform.  The corresponding state in the $SU(N)
\times SU(N)$ field theory is ${\cal B}_1$ of \eno{BaryonTwo}.  In the
$SU(N) \times SU(N+1)$ theory, one has instead
  \eqn{TwoBaryons}{
   \epsilon_{\alpha_1 \ldots \alpha_N} \epsilon^{\beta_1 \ldots \beta_{N+1}} 
    A^{\alpha_1}_{\beta_1} \ldots A^{\alpha_N}_{\beta_N} 
      \qquad \hbox{or} \qquad
   \epsilon_{\alpha_1 \ldots \alpha_N} \epsilon^{\beta_1 \ldots \beta_{N+1}} 
    A^{\alpha_1}_{\beta_1} \ldots A^{\alpha_N}_{\beta_N} 
     A^{\alpha_{N+1}}_{\beta_{N+1}}
  }
 where we have omitted $SU(2)$ indices.  Either the upper index
$\beta_{N+1}$, indicating a fundamental of $SU(N+1)$, or the upper
index $\alpha_{N+1}$, indicating a fundamental of $SU(N)$, is free.

How can this be in supergravity?  The answer is simple: the wrapped
D3-brane must have a string attached to it.  In the $\S^5/\Z_2$ theory
from which our original $SU(N) \times SU(N)$ theory descends via RG
flow, it is clear that a string ending on the holographic world-volume
transforms in the $(\N,1) \oplus (1,\N)$ of the gauge group.  The same
then should be true of the $T^{1,1}$ theory.  The new feature of the
domain wall is that a string must stretch from it to the wrapped
D3-brane.  There are two roughly equivalent ways to see that this
string must be present.  Most directly, one can recall that a D3-brane
crossing a D5-brane completely orthogonal to it leads to the
production of a string stretched between the two.  In flat space this
effect was discussed in detail in \cite{bdg,dfk} and is U-dual to the
brane creation process discovered in \cite{HW}.\footnote{Note that
while the D5-brane and the D3-brane may not be strictly orthogonal in
our setup, they do traverse complementary directions in $T^{1,1}$, so
together they fill out eight spatial dimensions.  This is sufficient
for the arguments of \cite{HW,bdg,dfk} to apply.}  Equivalently, one
can proceed along the lines of \cite{Ed}, noting first that there is a
discontinuity of $\int_{\S^3} H_{RR}$ across a D5-brane.  Since we have
assumed that on one side the theory is the original $SU(N) \times
SU(N)$ theory, all the $H_{RR}$-flux should be through three-spheres
on the other side.  More precisely, on the $SU(N) \times SU(N+1)$
side, $H_{RR}$ is an element of the third cohomology group
$H^3(T^{1,1})$, which is one-dimensional.  
Using the basis one-forms generated by the vielbeins of
$T^{1,1}$, 
  \eqn{TViel}{\displaylines{
   e^\psi = \tf{1}{3} \left( d\psi + \cos\theta_1 \phi_1 + 
    \cos\theta_2 \phi_2 \right)  \cr
   e^{\theta_1} = \tf{1}{\sqrt{6}} d\theta_1 \qquad
   e^{\phi_1} = \tf{1}{\sqrt{6}} \sin\theta_1 d\phi_1  \cr
   e^{\theta_2} = \tf{1}{\sqrt{6}} d\theta_2 \qquad
   e^{\phi_2} = \tf{1}{\sqrt{6}} \sin\theta_2 d\phi_2 \ ,
  }}
 we can express the harmonic representatives of the second and third
cohomology groups as 
  \eqn{HReps}{\eqalign{
   e^{\theta_1} \wedge e^{\phi_1} - e^{\theta_2} \wedge e^{\phi_2}
    &\in H^2(T^{1,1})  \cr
   e^\psi \wedge e^{\theta_1} \wedge e^{\phi_1} - 
    e^\psi \wedge e^{\theta_2} \wedge e^{\phi_2} &\in H^3(T^{1,1}) \ .
  }}
 The D3-brane wrapping $\S^3_{(1)}$ needs a fundamental string attached
to it to compensate for the flux of $H_{RR}$ from the D5: $B_{RR}
\to B_{RR} - \tilde{f}$ on the D3-brane where $d \tilde{f} = 2 \pi
\delta_P$ and $\tilde{f}$ is the dual of the $U(1)$ world-volume field
strength on the D3-brane.

The baryon ${\cal B}_2$ corresponding to a D3-brane wrapped around the
three-sphere which is the orbit of the other $SU(2)$, $\S^3_{(2)}$,
also becomes a non-singlet in the $SU(N) \times SU(N+1)$ theory: it
transforms in the $(\overline{\N},1) \oplus (1,\overline{\N+1})$.  This is
appropriate because $\S^3_{(2)}$ is opposite $\S^3_{(1)}$ in homology,
as one can see from the minus sign in \HReps.  Thus the three-form
flux through the D3-brane changes sign, and the fundamental string
that runs from it to the domain wall must be of opposite orientation
to the previous case.  In effect, a D3-brane around $\S^3_{(2)}$ is
topologically equivalent to an anti-D3-brane around $\S^3_{(1)}$.

It may be objected at this point that nothing selects which gauge
group gets changed as one crosses a D5 domain wall.  There is no
problem here because in fact nothing in the original $T^{1,1}$
solution distinguishes the two gauge groups.  Our only claim is that
crossing a domain wall increments (or for anti-D5's, decrements) the
rank of one gauge group: we do not attempt to distinguish between
$SU(N) \times SU(N+1)$ and $SU(N+1) \times SU(N)$, if indeed there is
any difference other than pure convention.

The domain wall in $AdS_5$ made out of $M$ wrapped D5-branes has the
following structure: on one side of it the 3-form field $H_{RR}$
vanishes, while on the other side there are $M$ units of flux of
$H_{RR}$ through the $\S^3$.  Thus, the supergravity dual of the
$SU(N)\times SU(N+M)$ theory involves adding $M$ units of RR 3-form
flux to the $AdS_5\times T^{1,1}$ background. If $M$ is held fixed
while $N\to \infty$ then the additional 3-form field will not alter
the gravity and the 5-form background.  In particular, the presence of
the $AdS_5$ factor signals that the theory remains conformal to
leading order in $N$. This agrees with the fact that assigning
R-charge $1/2$ to the bifundamental fields $A_k$ and $B_l$ guarantees
that the beta functions for both $SU(N)$ and $SU(N+M)$ factors vanish
to leading order in $N$ (for $M\neq 0$ there are, however, $1/N$
corrections to the beta functions).

A different situation occurs if the large $N$ limit is taken with
fixed $M/N$. Then it is obvious that addition of $M$ flux quanta of
$H_{RR}$ will have back-reaction on the geometry even at leading order
in $N$. Some solutions with both 5-form and 3-form field strengths
were discussed in \cite{Romans}, but they were found to break all
supersymmetry. For comparison with ${\cal N}=1$ supersymmetric field
theory one presumably needs to find a static supergravity background
with the same degree of supersymmetry. We leave the search for such
backgrounds as a problem for the future.  Some interesting physics
motivates this search: for $N_1\neq N_2$ it is impossible to choose
the R-charges so that the beta functions for both $SU(N_1)$ and
$SU(N_2)$ vanish.  Correspondingly, in supergravity, the presence of
three-form flux generically necessitates a dilaton profile (and even
the converse is true for static, supersymmetric, bosonic, type~IIB
backgrounds).  Furthermore, the quartic superpotential $W=\lambda\Tr
(A_1B_1A_2B_2-A_1B_2A_2B_1)$ is no longer marginal.\footnote{We thank
M.~Strassler for pointing this out to us.}  Therefore, the
corresponding supergravity background is not expected to have the
$AdS_5\times X_5$ structure.

The $N_1 \neq N_2$ field theories are somewhat analogous to the
Standard Model, where $SU(2)$ and $SU(3)$ have positive and negative
beta functions, respectively, and left-handed quarks form
bifundamentals under these gauge groups.  Two salient differences
between the Standard Model and our theories are the chiral coupling of
the weak interactions and the presence of matter fields (leptons)
which are neutral under the larger gauge group.  An analysis along the
lines of \cite{seiberg} may help elucidate the possible ${\cal N}=1$
field theories.  Supergravity compactifications which have both 5-form
and 3-form fields turned on and which preserve ${\cal N}=1$
supersymmetry appear to be good candidates for their dual description.

\section{Other wrapped branes}
\label{Other}

In this section we list other admissible ways of wrapping
branes over cycles of $T^{1,1}$ and discuss their field theory
interpretation. Our discussion is quite analogous to that
given by Witten for $AdS_5\times \RP^5$ \cite{Ed}.

Since $\pi_1( T^{1,1} )$ is trivial, there are no states associated
with wrapping the 1-branes. For D3-branes there are two types of 
wrapping. One of them, discussed in the
previous section, involves 3-cycles and produces particles on
$AdS_5$ related to the baryon-like operators (\ref{BaryonOne}) and
(\ref{BaryonTwo}). The other involves wrapping a D3-brane over an
$\S^2$ and produces a string in $AdS_5$.
The tension of such a ``fat'' string
scales as 
$L^2/\kappa \sim N (g_s N)^{-1/2}/\alpha'$. The non-trivial
dependence of the tension on the 't~Hooft coupling $g_s N$ indicates
that such a string is not a BPS saturated object. This should be
contrasted with the tension of a BPS string obtained by wrapping a
D5-brane over $\RP^4$ in \cite{Ed}, which is $\sim N/\alpha'$.

In discussing wrapped 5-branes, we will limit explicit statements
to D5-branes: since a $(p,q)$ 5-brane is an $SL(2,\Z)$
transform of a D5-brane, our discussion may be immediately
generalized to wrapped $(p,q)$ 5-branes using the $SL(2,\Z)$
symmetry of the Type IIB string theory.
If a D5-brane is wrapped over the entire $T^{1,1}$ then, according to
the arguments in \cite{Ed,GO}, it serves as a vertex connecting $N$
fundamental strings. Since each string ends on a charge in the
fundamental representation of one of the $SU(N)$'s, the resulting
field theory state is a baryon built out of external quarks.  
A D5-brane wrapped over an ${\bf S}^2$ produces a domain wall
discussed in the previous section.

If a
D5-brane is wrapped over an ${\bf S}^3$ then we find a membrane in
$AdS_5$. Although we have not succeeded in finding its field theoretic
interpretation, let us point out the following interesting effect.
Consider positioning a ``fat'' string made of a wrapped D3-brane
orthogonally to the membrane. As the string is brought through the
membrane, a fundamental string 
stretched between the ``fat'' string and the membrane
is created. The origin of this effect
is, once again, creation of fundamental strings by crossing D5
and D3 branes, as discussed in \cite{bdg,dfk}.

\section{Conclusions}
\label{Conclude}

The $AdS_5\times T^{1,1}$ model of \cite{KW} is the first example of a
supersymmetric holographic theory based on a compact manifold which is not
locally $\S^5$.  Correspondingly, the quantum field theory description
in terms of an ${\cal N}=1$ $SU(N) \times SU(N)$ gauge theory is in no
way a projection of the ${\cal N}=4$ theory.  

In the context of this model, we have provided a string theory
description of baryon-like operators formed from a symmetric product
of $N$ bifundamental matter fields, fully antisymmetrized on upper and
lower color indices separately.  The dual representation of
such an operator is a D3-brane wrapped around an $\S^3$ embedded in
$T^{1,1}$.  Two natural ways of embedding $\S^3$ are as orbits of
either of the two $SU(2)$ global symmetry groups of the theory.  A
D3-brane wrapping an orbit of one $SU(2)$ can be regarded classically
as a charged particle allowed to move on the $\S^2$ which parametrizes
the inequivalent orbits.  The five-form flux supporting the $AdS_5
\times T^{1,1}$ geometry acts as a magnetic field through this $\S^2$,
and the quantum mechanical ground states fill out an $N+1$-dimensional
representation of the other $SU(2)$.  All this meshes beautifully with
the field theory because the $N$ matter fields are doublets of the
$SU(2)$'s.  Moreover, the 3-volume of the $SU(2)$ orbits gives a
dimension for the operators, $3N/4$, which is precisely matched by the
field theory.

We have used this baryon construction to argue that D5-branes wrapped
around the 2-cycle of $T^{1,1}$ act as domain walls separating the
original $SU(N) \times SU(N)$ theory from a $SU(N) \times SU(N+1)$
theory.  The essential point is that crossing a wrapped D3-brane
through a D5-brane creates a string stretched between the two, so that
the baryon is no longer a singlet, but rather a fundamental of one of
the gauge groups.  This tallies with the field theory, because when
one attempts to antisymmetrize the color indices on a product of $N$
or $N+1$ bifundamentals of $SU(N) \times SU(N+1)$, one is always left
with one free index.  Our treatment of the domain walls has been
restricted to the test brane approximation.  Further evidence for the
fact that the D5-brane domain walls lead to 
$SU(N_1) \times SU(N_2)$ gauge theories, 
as well as perhaps some new phenomena,
may arise when one understands the full supergravity solution.

\section*{Acknowledgements}
 We thank D.-E.~Diaconescu, 
N.~Seiberg, M.~Strassler, A.~Strominger, W.~Taylor, and especially 
O.~Aharony and E.~Witten, for valuable discussions. We
are grateful to the Institute for Advanced Study, where this work
was initiated, for
hospitality.  This work was supported in part by the NSF grant
PHY-9802484, the US Department of Energy grant DE-FG02-91ER40671 and
by the James S. McDonnell Foundation Grant No. 91-48.


\end{document}